\documentclass[preprint,12pt]{elsarticle}	
\usepackage{amsmath,amsfonts,amssymb,bm}
\usepackage{algorithm}
\usepackage{algpseudocode}
\usepackage{xcolor}
\journal{Computers and Structures}
\begin{document}
\begin{frontmatter}
	\title{On the similarity of meshless discretizations of Peridynamics and Smooth-Particle Hydrodynamics}
	\author{G.~C.~Ganzenm\"uller, S. Hiermaier, and M. May}

	\address{Fraunhofer Ernst-Mach Institute for High-Speed Dynamics, EMI, Eckerstr. 4, D-79104 Freiburg i. Br., Germany} 
	\begin{abstract}
	{\color{black}
	This paper discusses the similarity of meshless discretizations of Peridynamics and
Smooth-Particle-Hydrodynamics (SPH), if Peridynamics is applied to classical material models based
on the deformation gradient. We show that the discretized equations of both methods coincide if
nodal integration is used. This equivalence implies that Peridynamics reduces to an old meshless
method and all instability problems of collocation-type particle methods apply. These instabilities
arise as a consequence of the nodal integration scheme, which causes rank-deficiency and leads to
spurious zero-energy modes. As a result of the demonstrated equivalence to SPH, enhanced
implementations of Peridynamics should employ more accurate integration schemes.}
  	\end{abstract}

	\begin{keyword}
		meshless methods \sep Peridynamics \sep Smooth-Particle Hydrodynamics
	\end{keyword}
\end{frontmatter}

\section{Introduction}

The Peridynamic theory, originally devised by Silling \cite{Silling:2000/a}, is a nonlocal extension
of classical continuum mechanics, which is based on partial differential equations. Since partial
derivatives do not exist on crack surfaces and other singularities, the classical equations of
continuum mechanics cannot be applied directly when such features are present. In contrast, the
Peridynamic balance of linear momentum is formulated as an integral equation, which remains valid in
the presence of material discontinuities. Therefore, the Peridynamic theory can be applied directly
to modelling both bulk \textit{and} interface properties, using the same mathematical model.
Additionally, Peridynamics is readily implemented in a meshless formulation, which facilitates the
simulation of large deformations when compared to the traditional mesh-based Finite-Element method
used for simulating solid mechanics in the classical continuum theory \cite{Bower:2011/a}.

With these desirable features, Peridynamics has received considerable attention by researchers
interested in numerically describing fundamental crack growth and failure effects in brittle
materials \cite{Silling:2010/a, Ha:2010/a, Ha:2011/a, Agwai:2011/a}.  However, the scope of the
original Peridynamic formulation included only so-called \textit{bond-based} models
\cite{Silling:2005/a} which were limited to a fixed Poisson ratio for linear isotropic materials and
could not describe true plastic yielding. Further development of the theory led to
\textit{state-based} models \cite{Silling:2007a}, which can, in principle, describe all classic
material behaviour. In particular, the state-based theory offers a route to approximating the
classical deformation gradient, which can be used to obtain a classical stress tensor. The stress
tensor can then be converted into nodal forces using a Peridynamic integral equation. This promising
situation, i.e., the ability to use classical material models with a method that remains valid at
material discontinuities, has prompted a number of studies where plastic yielding, damage, and
failure was simulated using this new meshless method \cite{Warren:2009/a, Foster:2010/a}.

However, there exists a demand for studies which compare Peridynamics to other meshless methods.
While the theoretical correspondence of Peridynamics with classical elasticity theory has been
established \cite{Silling:2008/a}, no information is available on the accuracy of the discretized
Peridynamic expression suitable for computer implementation.  Relevant questions that need to be
addressed include, but are not limit to, whether linear fields can be exactly reproduced by the
discretized theory, and what the the order of convergence is when Peridynamic solutions are compared
against exact results. Little is known about how common problems encountered with other meshless
methods, e.g., the tensile instability \cite{Swegle:1995/a} or the rank deficiency problem
\cite{Belytschko:2000/a} affect Peridynamics. {\color{black}As an exception to this general observation, Bessa
\textit{et al} \cite{Bessa:2014/a} have published a study which demonstrates the equivalence of
state-based Peridynamics and the reproducing Kernel Particle Method (RKPM), if nodal integration is
used. However, no published studies are available to the best of this author's knowledge, which
compare Peridynamics to other meshless methods.}

The goal of our paper is to elucidate on some properties of Peridynamics with respect to other
meshless methods. In particular, we establish that the discrete equations of the Peridynamic
formulation using classical material models is identical to a very well-known meshless technique:
Smooth Particle Hydrodynamics (SPH) in the Total-Lagrangian formulation. {\color{black} This equivalence
facilitates understanding of Peridynamics using the large body of literature already published for
other meshless methods, see e.g. \cite{Nguyen:2008/a} and \cite{Rabczuk:2010/a}. The key
observations of our work address two issues: (i) Discretizations of Peridynamics directly arrive at
correct equations which conserve linear and angular momentum. These features can only be obtained
in SPH by assuming \textit{ad hoc} corrections such as explicict symmetrization. (ii) All of the
problems that apply to collocation-type particle methods also apply to Peridynamics if this theory
is discretized using nodal integration.}

The remainder of this paper is organized as follows: we begin by deriving the fundamental expressions
of the SPH approximation including the most important corrections for this method, which allow it to
be used with the minimal level of accuracy required for solid mechanics simulations. Then, the
essential Peridynamic expressions required for simulating classical material models are derived.
Building on this foundation, the equivalence of SPH and this particular variant of Peridynamics is
shown. Finally, the implications of this observation are discussed and Peridynamics (with classical
material models) is characterized using the established terminology encountered in the SPH literature.


\section{Total Lagrangian SPH}
Smooth Particle Hydrodynamics \cite{Lucy:1977/a} was originally devised as a Lagrangian particle method with
the smoothing kernel moving with the particle, thus redefining the interaction neighbourhood for
every new position the particle attains. In this sense, the kernel of the original SPH formulation
has Eulerian character, as other particles move through the interaction neighbourhood. The tensile
instability \cite{Swegle:1995/a} encountered in SPH, where particles clump together under negative pressure conditions,
has been found to be caused by the Eulerian kernel functions \cite{Belytschko:2000/a}. Consequently,
Total Lagrangian formulations were developed \cite{Bonet:2002/a, Rabczuk:2004a, Vignjevic:2006a},
which use a constant reference configuration for defining the interaction neighborhood of
the particles.  see Typically, the initial, undeformed configuration is taken for this purpose. In the
following, this concept and the associated nomenclature is briefly explained with the limited scope
of obtaining SPH expressions that are to be compared with the Peridynamic expressions. For a more
detailed derivation, the reader is referred to the works cited above.

\subsection{Total Lagrangian formulation}
In the total Lagrangian formulation, conservation and constitutive equations are expressed
in terms of the material coordinates $\bm{X}$, which are taken to be the coordinates of the initial,
undeformed reference configuration. A mapping between the current coordinates, and the reference
coordinates describes the body motion at time $t$:
\begin{equation}
	\bm{x} = \phi(\bm{X}, t),
\end{equation}
Here, $\bm{x}$ are the current, deformed coordinates and $\bm{X}$ the reference (Lagrangian) coordinates. The
displacement $\bm{u}$ is given by
\begin{equation}
	\bm{u} =\bm{x} - \bm{X},
\end{equation}
The conservation equations for mass, impulse, and energy in the total Lagrangian formulation are given by
\begin{eqnarray}
	\rho J = \rho_0 J_0\\
	\ddot{u} = \frac{1}{\rho_0} \bm{\nabla}_0 \cdot \bm{P} \label{eqn:divergence}\\
	\dot{e} = \frac{1}{\rho_0} \dot{\bm{F}}:\bm{P}^{T},
\end{eqnarray}
where $J$ and $J_0$ are the current and initial Jacobian determinants. $\rho$ is the
current mass density and $\rho_0$ is the initial mass density, $\bm{P}$ is the nominal stress tensor
(the transpose of the first Piola-Kirchhoff stress tensor), $e$ is the internal energy,
$\bm{\nabla}_0$ is the gradient or divergence operator expressed in the material coordinates, and
$\bm{F}$ denotes the deformation gradient,
\begin{equation} \label{eqn:defgrad}
	\bm{F} = \frac{\mathrm{d}\bm{x}} {\mathrm{d}\bm{X}} = \frac{\mathrm{d}\bm{u}}
{\mathrm{d}\bm{X}} + \bm{I},
\end{equation}

\subsection {The SPH Approximation}
The SPH approximation for a scalar function $f$ in terms of the Lagrangian coordinates can be
written as
\begin{equation}
	f(\bm{X_i}) = \sum \limits_{j \in \mathcal{S}} V_j^0 f(\bm{X_j}) W_i \left( X_{ij} \right)
\end{equation}
The sum extends over all particles within the range of a scalar weight function $W_i$, which is
centered at position $\bm{X_i}$ and depends on the distance between coordinates $\bm{X}_i$ and
$\bm{X}_j$, $X_{ij}=\lVert \bm{X_j} - \bm{X_i} \rVert$. $V^0$ is the volume associated with a
particle in the reference configuration.  The weight function is typically chosen to be radially
symmetric and have compact support, i.e., it includes only neighbors within a certain radial
distance. This domain of influence is denoted $S$.

The SPH approximation of a derivative of $f$ is obtained by operating directly with the gradient operator on the kernel functions,
\begin{equation} \label{eqn:sph_derivative_approximation}
	\bm{\nabla} f(\bm{X_i}) = \sum \limits_{j \in \mathcal{S}} V_j^0 f(\bm{X_j}) \bm{\nabla} W_i \left(
X_{ij} \right),
\end{equation}
where the gradient of the kernel function is defined as follows:
\begin{equation}
	\bm{\nabla} W_i(X_{ij}) = \left ( \frac{\mathrm{d}W(X_{ij})}{\mathrm{d}X_{ij}} \right ) \frac{\bm{X}_j - \bm{X}_i}
{X_{ij}}
\end{equation}

The conditions for the zeroth- and first-order completeness of the SPH approximation are stated as
follows:
\begin{eqnarray}
	\sum \limits_{j \in \mathcal{S}} V_j^0 W_i \left( X_{ij} \right) = 1
\label{eqn:zeroth_order_completeness}\\
	\sum \limits_{j \in \mathcal{S}} V_j^0 \bm{\nabla} W_i \left( X_{ij} \right) = 0 \label{eqn:first_order_completeness}
\end{eqnarray}
In the simple form as stated here, neither of the completeness conditions are fulfilled by the SPH
approximation. An \textit{ad-hoc} improvement consists in adding
eqn.~(\ref{eqn:first_order_completeness}) to eqn.~\ref{eqn:sph_derivative_approximation}, such that a
\textit{symmetrized} approximation for the derivative of a function is obtained,
\begin{equation} \label{eqn:sph_derivative_approximation_symmetrized}
	\bm{\nabla} f(\bm{X_i}) = \sum \limits_{j \in \mathcal{S}} V_j^0 \left( f(\bm{X_j}) - f(\bm{X_i})
\right ) \bm{\nabla} W_i \left( X_{ij} \right)
\end{equation}
The symmetrization does not result in first-order completeness, however, it yields zeroth-order
completeness for the derivatives of a function.

\subsection {Restoring First-Order Completeness}
In order to fulfill first-order completeness, the SPH approximation has to reproduce the constant
gradient of a linear field. A number of correction techniques \cite{Randles:1996a, Bonet:1999a,
Vignjevic:2000a} exploit this condition as the basis for correcting the gradient of the SPH weight
function,
\begin{equation} \label{eqn:first_order_completeness_vector}
	\sum \limits_{j \in \mathcal{S}} V_j^0 (\bm{X}_j - \bm{X_i}) \otimes \bm{\nabla} W_i(X_{ij}) \overset{!}{=} \bm{I},
\end{equation}
where $\bm{I}$ is the diagonal unit matrix. Based on this expression, a corrected kernel gradient can be
defined:
\begin{equation} \label{eqn:correct_SPH_gradient}
	\tilde{\bm{\nabla}} W_i(X_{ij}) = \bm{L}_i^{-1} \bm{\nabla} W_i(X_{ij}),
\end{equation}
which uses the correction matrix $\bm{L}$, defined as:
\begin{equation} \label{eqn:SPH_correction_matrix}
	\bm{L}_i = \sum \limits_{j \in \mathcal{S}} V_j^0 \bm{\nabla} W_i(X_{ij}) \otimes (\bm{X}_j -
	\bm{X_i}).
\end{equation}

By construction, the corrected kernel gradient now satisfies eqn.~(\ref{eqn:first_order_completeness_vector}),
\begin{equation} \label{eqn:first_order_completeness_vector)}
	\sum \limits_{j \in \mathcal{S}} V_j^0 (\bm{X}_j - \bm{X_i}) \otimes \bm{L}_i^{-1} \bm{\nabla}  W_i(X_{ij})
= \bm{I},
\end{equation}

\subsection{Corrected SPH expressions for Solid Mechanics}
For calculating the internal forces of a solid body subjected to a deformation, expressions are required
for (i) the deformation gradient, (ii) a constitutive equation which provides a stress tensor as
function of the deformation gradient, and (iii) an expression for transforming the stresses into
forces acting on the nodes which serve as the discrete representation of the body.

The deformation gradient is obtained by calculating the derivative of the displacement field, i.e.,
by using the symmetrized SPH derivative approximation,
eqn.~(\ref{eqn:sph_derivative_approximation_symmetrized}), for eqn.~(\ref{eqn:defgrad}):
\begin{equation} \label{eqn:SPH_defgrad}
	\bm{F}_i^{SPH} = \sum \limits_{j \in \mathcal{S}} V_j^0 (\bm{u}_j - \bm{u_i}) \otimes \bm{L}_i^{-1} \bm{\nabla}
W_i(X_{ij}) + \bm{I}.
\end{equation}
Note that in the above equation, the corrected kernel gradient has been introduced via the matrix
$\bm{L}_i^{-1}$. The SPH approximation of the stress divergence, eqn.~(\ref{eqn:divergence}), is not so clear.
Depending on how it is performed, several different approximations can be obtained
\cite{Gingold:1982a}. The most frequently used form, which is variationally consistent with respect
to an energy minimization principle \cite{Bonet:1999a}, is the following: 
\begin{equation} \label{eqn:SPH_force_uncorrected}
	\bm{f}_i = \sum \limits_{j \in \mathcal{S}} V_i^0 V_j^0 \left ( \bm{P}_j + \bm{P}_i \right )
	\bm{\nabla} W_i(X_{ij}).
\end{equation}
For a radially symmetric kernel which depends only on distance, the antisymmetry property $\bm{\nabla} W_i(X_{ij}) = -
\bm{\nabla} W_j(X_{ji})$ holds. Therefore, the above force expression will conserve linear momentum
exactly, as $\bm{f}_{ij} = - \bm{f}_{ji}$. We use the antisymmetry property of the kernel gradient to rewrite the force
approximation as follows:
\begin{eqnarray} \label{}
	\bm{f}_i & = & \sum \limits_{j \in \mathcal{S}} V_i^0 V_j^0 \left ( \bm{P}_i \bm{\nabla} W_i(X_{ij}) + \bm{P}_j \bm{\nabla} W_i(X_{ij}) \right ) \\
                 & = & \sum \limits_{j \in \mathcal{S}} V_i^0 V_j^0 \left ( \bm{P}_i \bm{\nabla} W_i(X_{ij}) - \bm{P}_j \bm{\nabla} W_j(X_{ji}) \right ).
\end{eqnarray}
Replacing the uncorrected kernel gradient with the corrected gradient (c.f. eqn.~(\ref{eqn:correct_SPH_gradient}), the
following expression is obtained:
\begin{equation} \label{}
	\bm{f}_i = \sum \limits_{j \in \mathcal{S}} V_i^0 V_j^0 \left ( \bm{P}_i \bm{L}_i^{-1} \bm{\nabla} W_i(X_{ij}) - \bm{P}_j
\bm{L}_j^{-1} \bm{\nabla} W_j(X_{ji}) \right )
\end{equation}
This corrected force evaluation conserves linear momentum due to its antisymmetry with respect
to interchange of the particle indices $i$ and $j$, i.e., $\bm{f}_{ij} = - \bm{f}_{ji}$.  The here
constructed antisymmetric force expression is usually not seen in the literature. In contrast, it
seems to be customary \cite{Randles:1996a, Bonet:1999a, Vignjevic:2000a} to directly insert the
corrected kernel gradient into eqn.~(\ref{eqn:SPH_force_uncorrected}), which destroys the local
conservation of linear momentum. We note that the above construction of the SPH expression that
conserves linear momentum is arbitrary, and in similar spirit to the \textit{ad-hoc} symmetrization
procedure encountered in eqn.~(\ref{eqn:sph_derivative_approximation_symmetrized}).


\section{Peridynamics for classical material models}
This section provides a concise derivation of the Peridynamic approximation of the deformation
gradient and the forces acting on particles. In contrast to the preceding section, where this
quantities have been derived for SPH, we employ a different nomenclature here, which is consistent
with the most relevant Peridynamic literature. In Peridynamics, each particle defines the origin of an influence domain, termed neighborhood
$\mathcal{H}$, of radius $\delta$. Within this neighborhood, vectors $\bm{\xi}$ from the origin to any point in
$\mathcal{H}$ exist, see the following figure.

\begin{figure}[!ht]
\begin{center}
\includegraphics[width=0.75\textwidth]{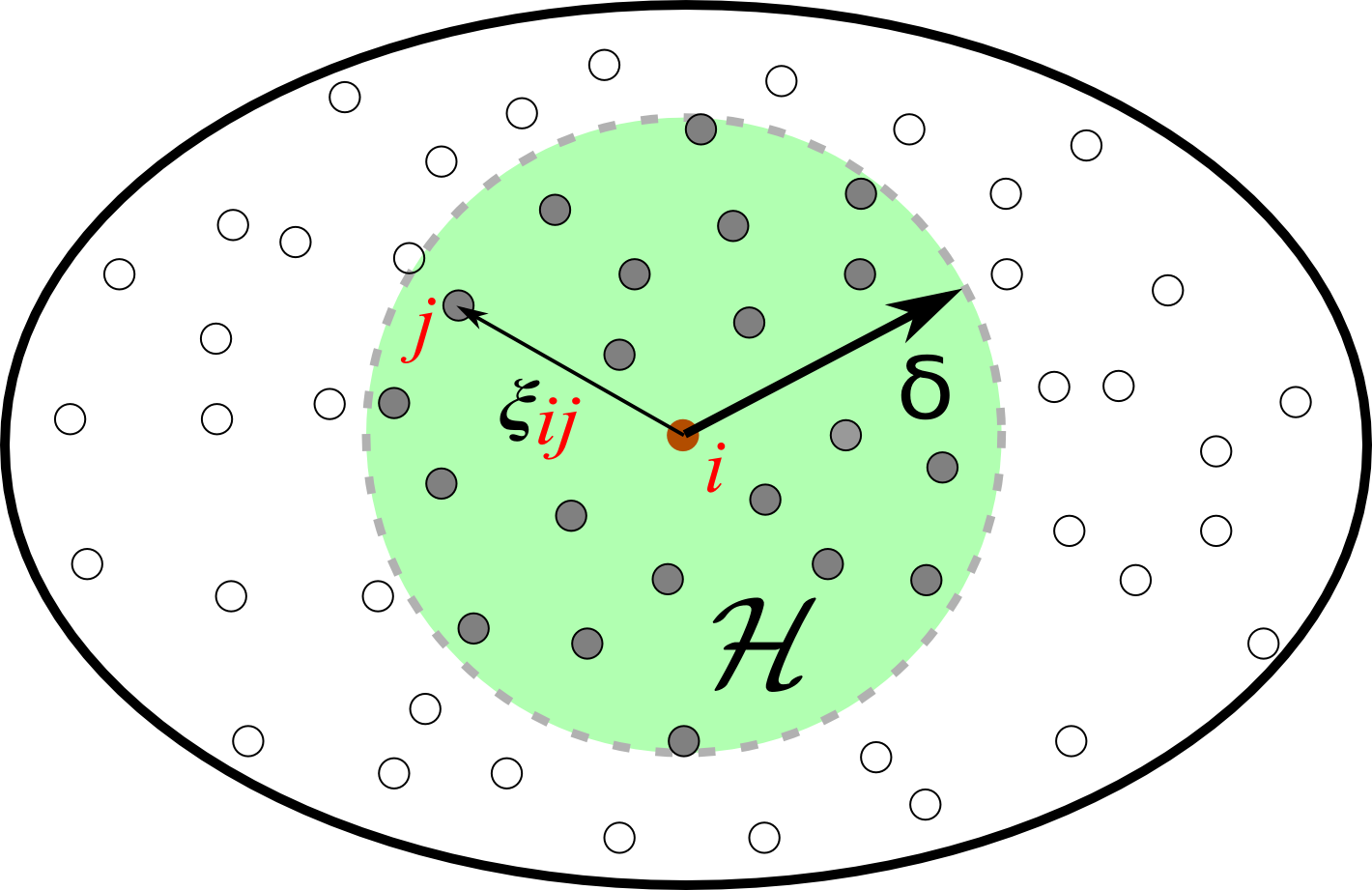}
\end{center}
\caption{
In the meshless Peridynamic method, particle interactions are defined in a reference configuration
which is shown here. A region of influence termed $\mathcal{H}$ is defined by a radial
cutoff $\delta$ around a particle $i$. Bonds exist between $i$ and all particles $j$ within
$\mathcal{H}$. These bonds are denoted here by $\bm{\xi}_{ij}$, of which only one is shown above.
Note that this representation already employs discrete particle locations whereas the Peridynamic
theory assumes a continuum of points.}
\label{fig:PD_peridynamics}
\end{figure}


\subsection{Vector states}
Central to the state-based Peridynamic theory is the concept of
\textit{states}. States are functions that act on on vectors. This is written
in the following way,
\begin{equation}
	\underline{\bm{A}} \left< \bm{\xi} \right> = \bm{\xi}',
\end{equation}
where the state $\underline{\bm{A}}$ has acted on the vector $\bm{\xi}$ to produce a different vector
$\bm{\xi}'$. Angular brackets indicate which vector the state acts upon.  The mapping results
produce tensors of different order, depending on the nature of the state. \textit{Vector states} of order one map a
vector to a different vector. \textit{Scalar states} of order zero produce a scalar for every vector
they act on. For the purpose of this paper we only need to be concerned with the following states:

\textit{The reference position vector state} $\underline{\bm{X}}$ returns the original bond vector.
\begin{equation}
	\underline{\bm{X}} \left< \bm{\xi} \right> = \bm{\xi} \in \mathbb{R}^3
\end{equation}

\textit{The deformation state} $\underline{\bm{Y}}$ returns the deformed image of the original
bond vector. Let the original vector go from $\bm{X}$ to $\bm{X'}$, i.e., $\bm{\xi} = \bm{X'} - \bm{X}$.
Upon deformation,  the coordinates of these two points change to $\bm{x}$ and $\bm{x'}$. The
deformation state vector returns the deformed image of the original vector, i.e.,
\begin{equation}
	\underline{\bm{Y}} \left< \bm{\xi} \right> = \bm{x'} - \bm{x}
\end{equation}

The \textit{influence function} $\underline{\omega}$ is a scalar state that returns a 
number $w$ that depends only on the magnitude of $\bm{\xi}$.
\begin{equation}
	\underline{\omega} \left< \xi \right> = w \in \mathbb{R}^+
\end{equation}

Vector and scalar states can be combined using the usual mathematical operations addition,
subtraction, multiplication, etc. The conventions are detailed in \cite{Silling:2007a}. Here, we
only need the definition of the \textit{tensor product} between two vector states $\underline{\bm{A}}$ and $\underline{\bm{B}}$,
\begin{equation}
	\underline{\bm{A}} * \underline{\bm{B}} = \int \limits_{\mathcal{H}} \underline{\omega} \left < \xi \right >
\underline{\bm{A}} \left < \bm{\xi} \right > \otimes \underline{\bm{B}} \left <
\bm{\xi} \right > \mathrm{d}V_{\bm{\xi}}
\end{equation}
This tensor product is required for the following Peridynamic concepts. Firstly, the \textit{shape
tensor} is defined:
\begin{equation}
	\mathbf{K} = \underline{\bm{X}} * \underline{\bm{X}} = \int \limits_{\mathcal{H}}
\underline{\omega} \left < \xi \right > \bm{\xi} \otimes \bm{\xi} \mathrm{d}V_{\bm{\xi}}
\end{equation}
$\bm{K}$ is symmetric and real and thus positive-definite, implying that it can be inverted.
The shape tensor is then used in the concept of \textit{tensor reduction}, which is an operation
that produces a second order tensor from two vector states:
\begin{equation}
	\mathbf{C} = \left( \underline{\bm{A}} * \underline{\bm{B}} \right) \bm{K}^{-1} = 
                     \left(\int \limits_{\mathcal{H}} \underline{\omega} \left < \xi \right >
                     \underline{\bm{A}} \left < \bm{\xi} \right >  \otimes
                     \underline{\bm{B}} \left < \bm{\xi} \right > \mathrm{d}V_{\bm{\xi}} \right)  \bm{K}^{-1}
\end{equation}


\subsection{Calculation of the deformation gradient tensor in Peridynamics}
The deformation gradient tensor may be approximated as a tensor reduction of deformation vector state and
reference position vector state according to eqn.~(134) in \cite{Silling:2007a}:
\begin{equation}
	\bm{F} \approx (\underline{\bm{Y}}*\underline{\bm{X}}) \bm{K}^{-1}
\end{equation}
The necessary Peridynamic calculus for evaluating this quantity has been presented above. We now
make the transition from the continuum representation to a discrete nodal, or
particle representation, converting integrals to sums. A piecewise constant discretization of the
integrals defined above yields the discrete shape tensor as
\begin{equation}
	\bm{K}_i = \sum \limits_{j \in \mathcal{H}_i} V_j^0 \underline{\omega} \left < \xi_{ij} \right>
       \bm{\xi}_{ij} \otimes  \bm{\xi}_{ij}
\end{equation}
Here, the sum includes all particles $j$ within the neighborhood $\mathcal{H}_i$ of $i$ and $V_j^0$
is the volume of particle $j$. The discrete expression for the approximate deformation gradient is obtained as:
\begin{equation}
	\bm{F}_i^{PD} = \left ( \sum \limits_{j \in \mathcal{H}_i} V_j^0 \underline{\omega} \left <\xi_{ij} \right >
        \bm{x}_{ij} \otimes  \bm{\xi}_{ij} \right) \bm{K}_i^{-1} =
        \left ( \sum \limits_{j \in \mathcal{H}_i} V_j^0 \underline{\omega} \left <\xi_{ij} \right >
        \bm{u}_{ij} \otimes  \bm{\xi}_{ij} \right) \bm{K}_i^{-1} + \bm{I}
\end{equation}

\subsection{The Peridynamic force expression obtained from a classical stress tensor}
The fundamental equation of Peridynamics is an integral equation of force densities (force per
volume) which ensures local conservation of impulse, see eqn.~(28). in \cite{Silling:2007a}:
\begin{equation}
	\bm{f}^{PD} = V_{\bm{X}} \int \limits_{\mathcal{H}_{X}} \left \{ \underline{\bm{T}}[\bm{X}] \left < \bm{X}' - \bm{X} \right
> - \underline{\bm{T}}[\bm{X}'] \left < \bm{X} - \bm{X'} \right> \right \} \mathrm{d}V_{X'}
\end{equation}
Here, $\bm{T}[\bm{X}]$ is a vector state, which operates on the bond $\bm{\xi} = \bm{X}' - \bm{X}$
in order to yield a contribution to the force density at $\bm{X}$. The antisymmetric counterpart
$\bm{T}[\bm{X}']$ ensures balance of linear momentum and $V_{\bm{X}}$ is the volume associated with the
coordinate $\bm{X}$.
The vector state $\bm{T}$ is related the classical stress tensor, which can be computed using the
deformation gradient and a classical material model. Specifically, this relation is stated as
follows, see eqn.~(142). in \cite{Silling:2007a}:
\begin{equation}
	\underline{\bm{T}}[\bm{X}] \left < \bm{\xi} \right > = \underline{\omega} \left < \xi \right
> \bm{P}_{\bm{X}} \bm{K}^{-1}_{\bm{X}} \bm{\xi}
\end{equation}
The stress tensor $\bm{P}$ above is the nominal stress tensor which applies here as the stress 
is expressed in the reference configuration. The dependence on $\bm{X}$ signifies that both stress
and shape tensor have to be evaluated at coordinate $\bm{X}$.  Converting to a discrete particle
expression, one obtains:
\begin{eqnarray}
	\bm{f}_i^{PD} & = & \sum \limits_{j \in \mathcal{H}_i} V_i^0 V_j^0 \left \{ \underline{\bm{T}}_i \left < \bm{\xi}_{ij}
\right > - \underline{\bm{T}}_j \left < \bm{\xi}_{ji} \right > \right \} \\
	              & = & \sum \limits_{j \in \mathcal{H}_i} V_i^0 V_j^0 \left \{ \underline{\omega} \left < \xi_{ij} \right >
                          \bm{P}_i \bm{K}_i^{-1} \bm{\xi}_{ij} - \underline{\omega} \left < \xi_{ji} \right > \bm{P}_j \bm{K}_j^{-1}
\bm{\xi}_{ji} \right \}
\end{eqnarray}

\section{The Correspondence between Peridynamic and SPH expressions for deformation gradient and
particle forces} \label{sec:prove_equivalence}

The discrete Peridynamic approximation of both the deformation gradient and the particle forces
arising from a classical stress tensor are equal to the Total-Lagrangian SPH expressions with
linear kernel gradient correction. In order to show this equivalence, we introduce some changes in nomenclature:
\begin{eqnarray}
	\mathcal{H}_i & = & S_i \\
	\bm{\xi}_{ij} & = & \bm{X}_{ij}\\
	\underline{\omega} \left < \xi_{ij} \right > & = & \frac{1}{X_{ij}}
\frac{\mathrm{d}W(X_{ij})} {\mathrm{d}X_{ij}}. \label{eqn:PD_SPH_weight_function}
\end{eqnarray}
Note that the postulated equivalence in the last line above states that the derivative of the SPH
weight function has to equal the Peridynamic weight function. However, no implications arise from
this requirement, as suitable weight functions can be chosen which fulfills this requirement. 

\subsection{Equality of shape tensor and first-order correction matrix}
With these changes, the Peridynamic shape tensor becomes equal to the correction matrix required for
first-order consistent SPH:
\begin{eqnarray}
	\bm{K}_i & = & \sum \limits_{j \in \mathcal{H}_i} V_j^0 \underline{\omega} \left < \xi_{ij} \right >
\bm{\xi}_{ij} \otimes  \bm{\xi}_{ij} \\
                 & = & \sum \limits_{j \in \mathcal{S}_i} V_j^0  \frac{1}{X_{ij}} \frac{\mathrm{d}W(X_{ij})} {\mathrm{d}X_{ij}} \bm{X}_{ij}
\otimes  \bm{X}_{ij} \\
		 & = & \sum \limits_{j \in \mathcal{S}_i} V_j^0 \bm{\nabla} W_i(X_{ij}) \otimes (\bm{X}_j -	\bm{X_i}) \\
		 & \therefore & \nonumber \\
	\bm{K}_i & = & \bm{L}_i
\end{eqnarray}

\subsection{Equality of the deformation gradient}
In a similar manner, the Peridynamic expression of the deformation gradient can be shown to be
equal to the SPH approximation:
\begin{eqnarray}
	\bm{F}_i^{PD} & = & \left ( \sum \limits_{j \in \mathcal{H}_i} V_j^0 \underline{\omega} \left < \xi_{ij} \right >
\bm{u}_{ij} \otimes \bm{\xi}_{ij} \right) \bm{K}_i^{-1} + \bm{I}\\
	              & = & \left ( \sum \limits_{j \in \mathcal{S}} V_j^0 \frac{1}{X_{ij}} \frac{\mathrm{d}W(X_{ij})} {\mathrm{d}X_{ij}}
                            \bm{u}_{ij} \otimes \bm{X}_{ij} \right) \bm{L}_i^{-1} +\bm{I}\\
	              & = & \sum \limits_{j \in \mathcal{S}} V_j^0  
                            \bm{u}_{ij} \otimes \left ( \bm{L}_i^{-1} \frac{1}{X_{ij}} \frac{\mathrm{d}W(X_{ij})}
                            {\mathrm{d}X_{ij}} \bm{X}_{ij} \right ) +\bm{I} \\
		 & \therefore & \nonumber \\
	\bm{F}_i^{PD} & = & \bm{F}_i^{SPH}
\end{eqnarray}
Thus, the Peridynamic concept of reduction leads to the approximation of a tensor field, which is
correct to first order.


\subsection{Equality of particle forces} 
Using the same rules for changing the Peridynamic notation into SPH notation, one obtains from the Peridynamic force
expression:
\begin{eqnarray}
	\bm{f}_i^{PD} & = & \sum \limits_{j \in \mathcal{H}} V_i^0 V_j^0 \left \{ \underline{\omega} \left < \xi_{ij}
                            \right > \bm{P}_i \bm{K}_i^{-1} \bm{\xi}_{ij} -
                            \underline{\omega} \left < \xi_{ji} \right > \bm{P}_j \bm{K}_j^{-1} \bm{\xi}_{ji} \right \} \\
	& = & \sum \limits_{j \in \mathcal{S}} V_i^0 V_j^0 \left \{ \frac{1}{X_{ij}} \frac{\mathrm{d}W(X_{ij})} {\mathrm{d}X_{ij}}
\bm{P}_i \bm{L}_i^{-1} \bm{X}_{ij} - \frac{1}{X_{ji}} \frac{\mathrm{d}W(X_{ji})} {\mathrm{d}X_{ji}} \bm{P}_j \bm{L}_j^{-1} \bm{X}_{ji}
\right \} \\
	& = & \sum \limits_{j \in \mathcal{S}} V_i^0 V_j^0 \left \{ 
\bm{P}_i \bm{L}_i^{-1} \bm{\nabla} W_{i}(X_{ij}) - \bm{P}_j \bm{L}_j^{-1} \bm{\nabla} W_{j}(X_{ji})
\right \} \\
	& \therefore & \nonumber\\
	\bm{f}_i^{PD} & = & \bm{f}_i^{SPH}
\end{eqnarray}

{\color{black}
\section{Implementation and numerical results}
\subsection{Implementation issues}
The equality of Total-Lagrangian SPH and Peridynamic for classical material models with nodal
integration raises some advantageous implementation issues. The first point concerns the weight
function and its derivative, which is usually attributed great influence in SPH. In contrast,
Peridynamic implementations typically use a constant unit weight function, i.e., $\xi_{ij} =
1 \, \forall \, j \, \in \mathcal{H}_i$. Using the Peridynamics to SPH translation,
eqn.~(\ref{eqn:PD_SPH_weight_function}), this corresponds to a constant gradient for the SPH scheme.
Such a constant implies that the associated weight function is apparently not a proper SPH weight
function, i.e., it is certainly not normalized: $\int_{S} 1 \mathrm{d}V \ne 1$. This apparent contradiction is solved by the first-order
correction scheme used here: The \textit{ansatz} used for deriving the corrected gradient,
eqn.~(\ref{eqn:correct_SPH_gradient}) does not depend on any normalization factor as the
multiplication the matrix inverse of $\bm{L}$ cancels any constant factor in the weight function. In
fact, it can be shown that this route to a corrected kernel gradient represents a Moving
Least Squares formulation with a linear basis \cite{Mueller:2004/a, Lancaster:1981/a}. It can
be concluded that the weight function (or indeed its derivative) can be chosen quite
arbitrarily within this scheme. For the purpose of illustrating an implementation of the first-order
corrected scheme, the uncorrected kernel gradient is simply defined as $W_{ij} \bm{X}_{ij}$.
Algorithm 1 outlines an efficient implementation which moves the multiplication with the shape matrix
outside of the inner loop (loop over neighbors $j$ of particle $i$), in order to save unnecessary calculations.

\begin{algorithm}
\caption{First-order corrected SPH or deformation-gradient based Peridynamics with nodal
integration. Note that the constitutive law used here is expressed using the Green-Lagrange strain
and the Lam\'e parameters $\lambda$ and $\mu$.} \label{euclid}
\begin{algorithmic}[1]
\ForAll{$i$}
	\State $\bm{F}_i \gets \bm{0}$ \Comment{zero deformation gradient}
	\State $\bm{L}_i \gets \bm{0}$ \Comment{zero shape matrix}
	\State $\bm{f}_i \gets \bm{0}$ \Comment{zero force}
	\\
	\ForAll{$j \in \mathcal{S}_i$} \Comment{first loop over pairs $(i,j)$}
		\State \Comment{note that $W_{ij} \bm{X}_{ij}$ replaces the usual SPH kernel gradient}
		\State $\bm{F}_i \gets \bm{F}_i + V_j^0 W_{ij} \bm{x}_{ij} \otimes  \bm{X}_{ij}$ \Comment{sum deformation gradient}
		\State $\bm{L}_i \gets \bm{L}_i + V_j^0 W_{ij} \bm{X}_{ij} \otimes  \bm{X}_{ij}$ \Comment{sum shape matrix}
	\EndFor
	\\
	\State $\bm{L}_i \gets \bm{L}_{i}^{-1}$ \Comment{invert shape matrix}
	\State $\bm{F}_i \gets \bm{F}_{i} \bm{L}_i$ \Comment{compute corrected deformation gradient}
	\\
	\State \Comment{compute stress using linear-elastic constitutive law}
	\State $\bm{E}_i \gets \frac{1}{2} \left(\bm{F}_i^T \bm{F}_i - \bm{I} \right)$ \Comment{Green-Lagrange strain}
	\State $\bm{S}_i \gets \lambda \mathrm{Tr}\{\bm{E}_i\} + 2 \mu \bm{E}_i$ \Comment{$2^{nd}$ Piola-Kirchhoff stress}
	\State $\bm{P}_i \gets \bm{F}_i \bm{S}_i$ \Comment{$1^{st}$ Piola-Kirchhoff stress}
	\State $\bm{P}_i \gets \bm{P}_i \bm{L}_i$ \Comment{multiply stress tensor with shape matrix}
	\\
	\ForAll{$j \in \mathcal{S}_i$} \Comment{second loop over pairs $(i,j)$}
		\State $\bm{f}_{i} \gets \bm{f}_i + V_i^0 V_j^0 W_{ij} \left( \bm{P}_i + \bm{P}_j \right) \bm{X}_{ij}$  \Comment{sum force}
	\EndFor

	\Comment{move particle $i$ according to time integration scheme}
\EndFor
\end{algorithmic}
\end{algorithm}

This algorithm exhibits two main features: A first loop over all pairs of interacting particles is
required to define the shape matrix and deformation gradient. Naive implementations might split up
this loop into two independent loops and first calculate only the shape matrix. However, as the
kernel gradient correction depends only on the shape matrix of particle $i$, c.f.
eqn.~(\ref{eqn:SPH_defgrad}), these operations can be combined and the correct deformation gradient
obtained by post-multiplying with the shape matrix. With the corrected deformation gradient at hand,
a stress tensor is computed using the constitutive law. Here, a finite-strain theory is employed,
which provides a linear relationship between strain and stress, based on the Lam\'e parameters
$\lambda$ and $\mu$. The stress measure obtained this way is the Second Piola-Kirchhoff stress,
which is applicable to the reference configuration. This stress measure is pushed forward to the
current configuration (First Piola-Kirchhoff stress), as forces are applied to the current particle
positions. The calculation of the forces requires a second loop over all pairs of interacting
particles. To avoid multiplication of the kernel gradients with the shape matrix inside the second
loop, the shape matrices are multiplied with the stress tensors before entering this loop. Once the
forces are computed, the particle positions can updated according to any chosen time-integration
scheme.

\subsection{Numerical examples}
The aim of this work is to show the equivalence of first-order corrected Total-Lagrangian SPH and
Peridynamics applied to classical material models in combination with nodal integration. While
the equivalence has been mathematically proven in sec.~(\ref{sec:prove_equivalence}), it is
also of interest to demonstrate the effects of the instability which emerges if nodal integrations
used. To this end, the implementation according to Algorithm 1, in combination with a unit weight
function and a leap-frog time integrator is used. For comparison purposes, reference solutions which
do not suffer from rank-deficiency due to nodal integration are shown which are obtained using a
meshfree code that employs a stabilization technique based on higher-order derivatives, similar to
\cite{Vidal:2007/a}.

\subsubsection{Patch test and stability}
The patch test can be regarded as the most basic test for a solid mechanics simulation code. A strip
of some material that is discretized using several elements (or SPH / Peridynamic particles) is
subjected to a uniform strain field and the stress is computed. Regardless of the discretization,
the stress should be uniform everywhere. SPH / Peridynamics with first-order corrected kernel
gradients is well-known to pass this test. Here, however, we are interested in the subsequent
time-integration following such an initial perturbation. To this end, a random initial particle
configuration is obtained by discretizing a square patch of edge length 1 m and uniform area mass
density $\rho=1 \mathrm{kg/m^2}$ into 444 quadliteral area elements using a stochastic algorithm
based on triangular Voronoi tesselation and recombination of the triangles into quadliterals.
Particle volumes and masses are obtained from the quadliteral area and the prescribed mass density.
The kernel radius is adjusted individually for each particle such that approximately 12 neighbors
are within the kernel range. While the patch test can be easily simulated with an ordered initial
configuration of particles, a non-uniform particle configuration poses extreme stability challenges.
Figure \ref{fig:quad_mesh} shows the initial particle configuration. The material model is chosen as
linear elastic with $E=1 \:\mathrm{Pa}$ and $\nu=0.3$. The patch is uniformly stretched by 10\%
along both axes. After this initial perturbation, the system is integrated in time using a standard
leap-frog algorithm with a CFL-stable timestep, resulting in a periodic contraction and expansion
mode. Figure \ref{fig:quad_patch_comparison} compares plain first-order corrected Total-Lagrangian
SPH (equivalent to Peridynamics with nodal integration) with a meshfree solution that is stabilized
against the effects of rank-deficiency. In the unstabilized system, particle motion is not coherent
and particle disorder is observed already after the first contraction. After eight contractions,
particle disorder is very pronounced and the motion of the patch is entirely dominated by numerical
artifacts. In contrast, the stabilized simulation exhibits a completely coherent particle motion
which can can be continued for hundreds of oscillations.

\begin{figure}[!ht]
\begin{center}
\includegraphics[width=0.5\textwidth]{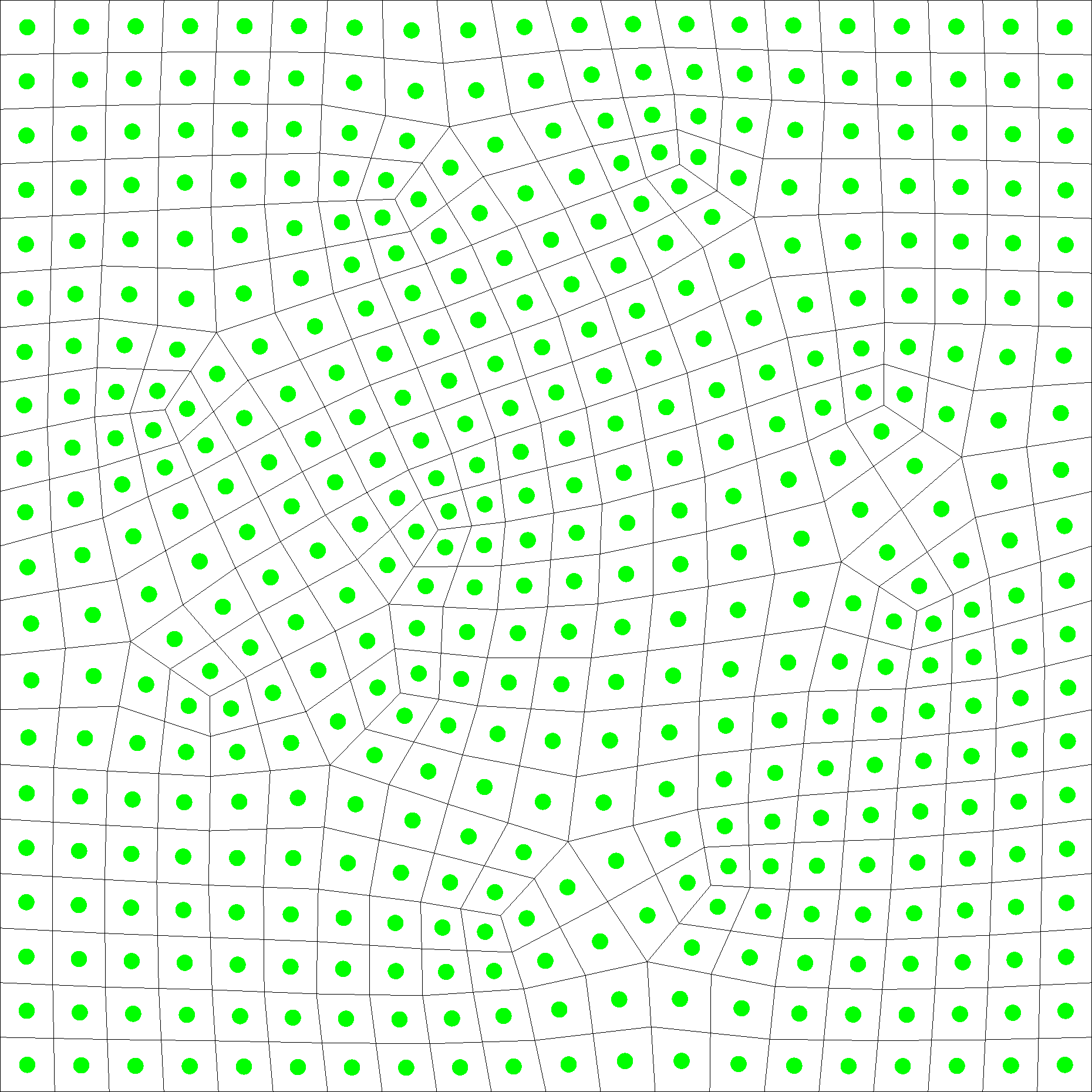}
\end{center}
\caption{Initial particle arrangement for the patch test stability example.
A random initial particle configuration is obtained by discretizing a square patch into 444 quadliteral area elements using a stochastic algorithm based on triangular Voronoi tesselation and recombination of the triangles into
quadliterals. The particle coordinates are taken to be the quadliteral element centers and particle
volumes are taken as the quadliteral area times a unit thickness.}
\label{fig:quad_mesh}
\end{figure}

\begin{figure}[!ht]
\begin{center}
\includegraphics[width=0.99\textwidth]{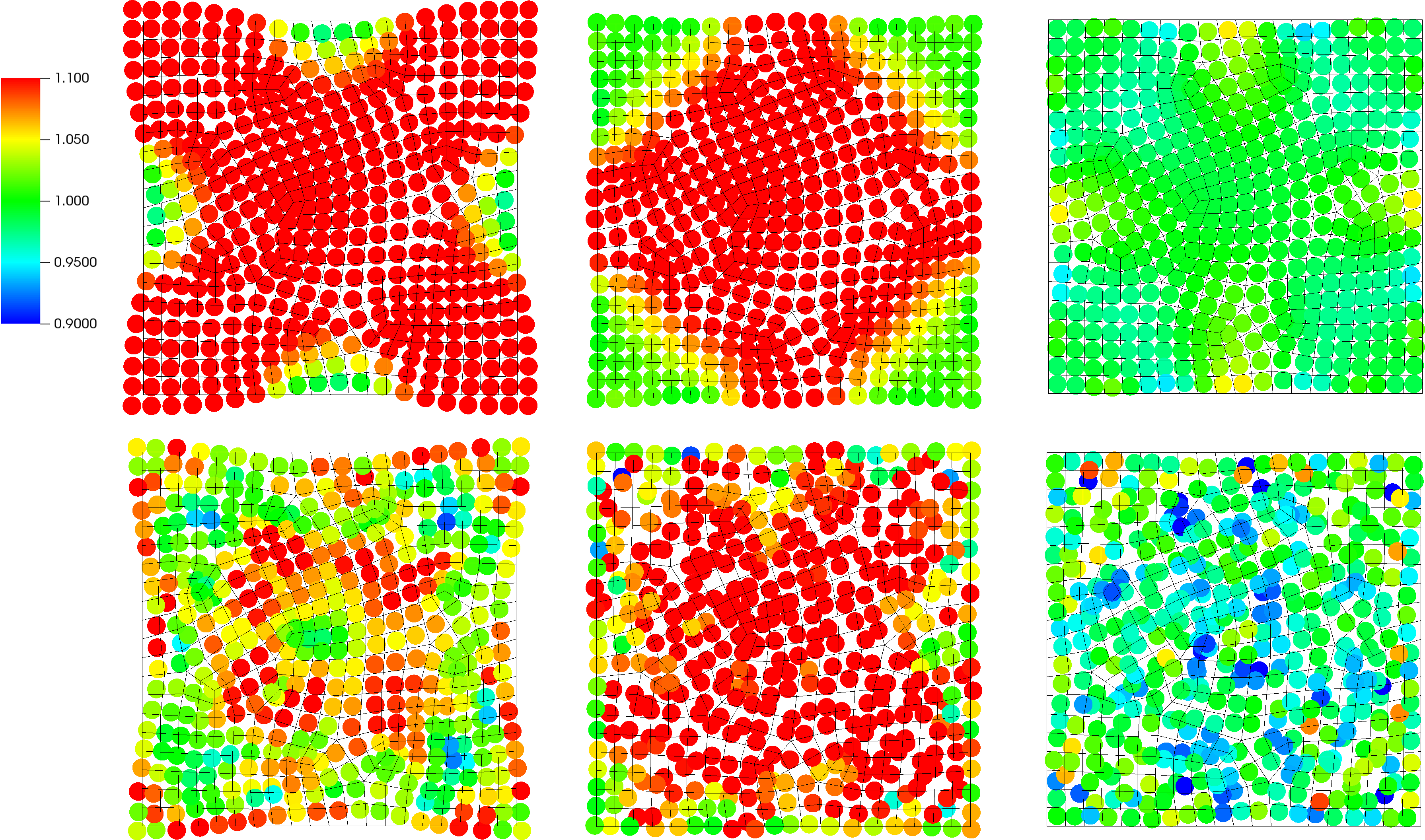}
\end{center}
\caption{Simulation snapshots comparing unstabilized and stabilized trajectories. A 2d patch of an
elastic material is perturbed by isotropic stretch of 10 \% and integrated in time, resulting in
a periodic contraction and expansion mode. The top row of images shows a stabilized
solution during the first, third, and tenth contraction
period. The bottom row shows corresponding snapshots from an unstabilized simulation. The mesh
in the background serves as a guide to the eye.}
\label{fig:quad_patch_comparison}
\end{figure}

\subsection{Bending of a beam}
This example considers a large deflection of a simply supported 2d beam with dimensions $8\,\mathrm{mm} \times
1\,\mathrm{mm}$, see Fig.~\ref{fig:beam_sketch}. The material model of the beam is taken as linear elastic with Youngs modulus $E=100\,\mathrm{GPa}$ and $\nu=0.3$.
A large boundary displacement of $u_y=8\,\mathrm{mm}$ is applied, and plain Total-Lagrangian SPH
results (equivalent to Peridynamics with nodal integration) are
qualitatively compared against a stabilized simulation code. Snapshots of the simulations corresponding to the same boundary
displacements, are shown in
Fig.~\ref{fig:beam_comparison}. In the first snapshot at $u_y=4.8\,\mathrm{mm}$, no obvious
difference between both simulations can be observed. For the second snapshot at
$u_y=6.6\,\mathrm{mm}$, the unstabilized simulation shows a pattern where pairs of particles in the
interior of the beam move together and do not follow the rotation of the beam's neutral line.
Clearly, the unstabilized system tries to minimize its elastic energy by assuming a configuration
which is not in agreement with a linear displacement field. This leads to an obviously wrong
particle configuration for the last snapshot at $u_y=8.0\,\mathrm{mm}$. In contrast, the stabilized
system shows particle displacements which appear very reasonable and exhibit no
instability.

\begin{figure}[!ht]
\begin{center}
\includegraphics[width=0.75\textwidth]{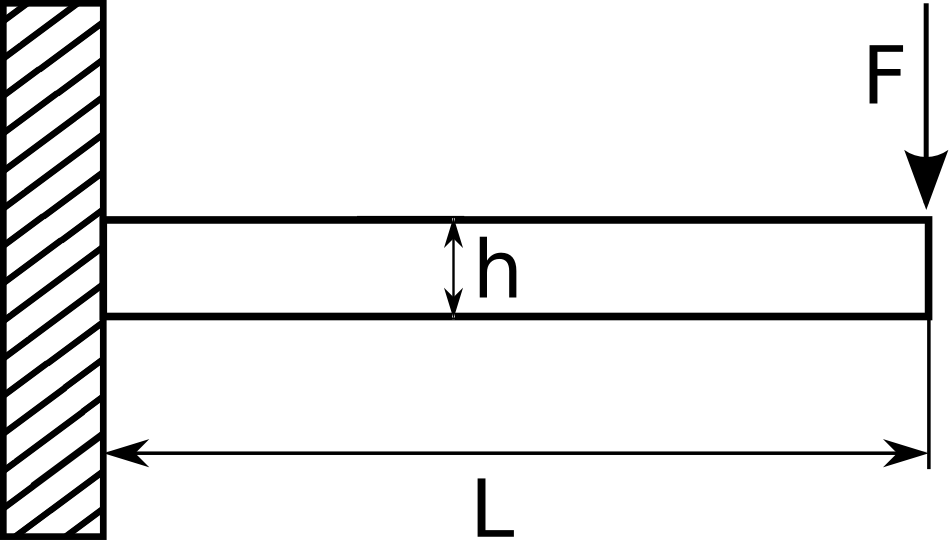}
\end{center}
\caption{Sketch of the simply supported beam with dimensions $L=8\,\mathrm{mm}$ and $h=1\, \mathrm{mm}$.}
\label{fig:beam_sketch}
\end{figure}

\begin{figure}[!ht]
\begin{center}
\includegraphics[width=0.99\textwidth]{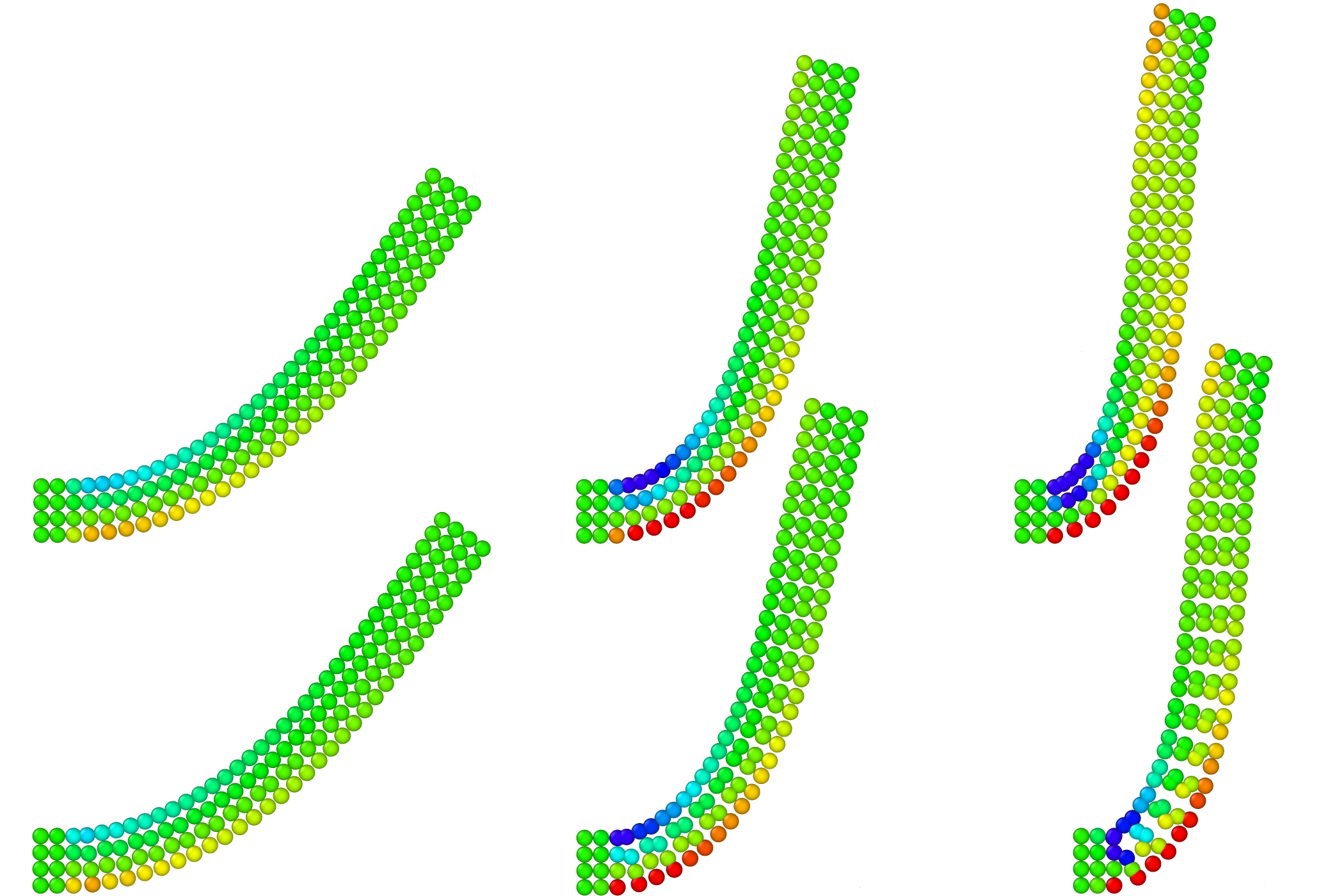}
\end{center}
\caption{Comparison of stabilized (top) and unstabilized results for a large deflection simulation
of the simply supported beam. The color coding represents the $xx$-component of the Second
Piola-Kirchhoff stress tensor, with red signifying tension and blue compression.}
\label{fig:beam_comparison}
\end{figure}
We conclude this test case by noting that Total-Lagrangian SPH, or -- equivalently -- Peridynamics
for classical material models based on the deformation gradient and with nodal integration, can become unstable for certain
loading conditions even with perfectly uniform reference particle configurations. Hence,
stabilization techniques, e.g. based on higher-order derivatives \cite{Vidal:2007/a} are required to
stabilize the solution.
}

\clearpage
\section{Discussion}
This work shows that Peridynamics -- when used with classical material models based on the deformation
gradient tensor and nodal integration as discretization technique -- is equivalent to Total
Lagrangian SPH with corrected gradients. This result allows to characterize this flavour of Peridynamics using the large
body of studies published for SPH. SPH is a collocation method with nodal integration that suffers
from two major problems: (i) the tensile instability, i.e., a numerical instability which results in
particle clumping under conditions of negative pressure, and (ii) susceptibility to zero-energy
modes which are caused by rank-deficiency due to nodal integration. Of these problems, the tensile
instability is resolved by the Total-Lagrangian formulation \cite{Belytschko:2000/a, Rabczuk:2004a},
and the rank-deficiency can be eliminated by using additional integration points
\cite{Randles:1996a}. Peridynamics with classical material models as presented here is a
Total-Lagrangian method and therefore shows no tensile instability.  However, if due to the use of
nodal integration, it is still a rank-deficient method.

The rank-deficiency does usually not manifest itself until the reference configuration is updated.
Such an update can be required as the Total-Lagrangian nature imposes restrictions on the magnitude
of deformations that can be handled with the approximate deformation gradient. Alternatively,
features of the material model, e.g. plasticity or failure, may require updates of the reference
configuration. These updates result in zero-energy modes which need to be treated with dissipative
mechanisms \cite{Vidal:2007/a}.

With the equivalence demonstrated in this work, Peridynamics applied to classical material models
and in combination with nodal integration does not result in a new numerical method, but is instead
a new derivation of an existing meshless method that is not free of problems. What are the
implications of these findings for the other Peridynamic theories? The bond-based theory
\cite{Silling:2005/a}, which can be interpreted as an upscaling of the established meshless method
Molecular Dynamics \cite{Seleson:2009/a}, does not require the calculation of a deformation gradient
and thus does not suffer from the associated problems due to rank-deficiency. The state-based
Peridynamic theory has also been formulated for some material models which are not based on the
classical deformation gradient. If these unconventional material models, e.g., the linear
Peridynamic solid \cite{Silling:2007a}, are discretized using nodal integration for computing
dilatation and shear strain, rank deficiency problems are also likely to emerge.

Recently, work relating the reproducing Kernel Particle Method (RKPM) to state-based Peridynamics
was published by Bessa \textit{et al} \cite{Bessa:2014/a}. In their work, the authors conclude that
state-based Peridynamics is equivalent to RKPM if nodal integration, i.e., collocation, is used.
These authors positively emphasize that the Peridynamic state-based approach is much faster compared
to RKPM using more elaborate (cell based) integration techniques. These results are also
understandable in light of this publication, because, it is known that RKPM with nodal integration
is equivalent to Moving-Least-Squares SPH (MLSPH), which is known to be equivalent to SPH with
corrected derivatives.  For a comprehensive review on this equivalence, which also addresses the
question as to how many really different meshless methods are known, see \cite{Vidal:2004/a}.
However, Bessa \textit{et al} \cite{Bessa:2014/a} disregard the unfortuitous implication of the
equivalence of Peridynamics, SPH, and RKPM with nodal integration: we have learned from a number of
SPH publications, that nodal integration causes problems, as it gives rise to zero-energy modes.
Therefore the use of nodal integration for Peridynamics is a bad thing to do.

It is worthwhile to emphasize that the mathematical foundation of Peridynamics is clear and
straightforward. Correct equations of motion emerge from this theory which conserve linear and
angular momentum, and approximate linear fields accurately. All of these desirable features can only
introduced into the SPH approximation by \textit{ad-hoc} procedures. e.g. explicit symmetrization.
Thus, if anything, the Peridynamic theory provides us with a better route to deriving meshless
discretizations than the SPH method. If the simplest form of meshless discretization, namely nodal
integration, is used, the advantages of Peridynamics over SPH vanish and the discrete expressions of
both methods become equal. Stabilization schemes which address the rank-deficiency problem such as
the scheme due to Vidal \textit{et al.} \cite{Vidal:2007/a} then need to be employed in order to keep
the simulation stable. Future work should therefore address enhanced integration schemes for the
Peridynamic theory.

\end{document}